\documentclass[a4paper,11pt]{article}
\pdfoutput=1 

\usepackage{jinstpub}

\title{ CaloCube: a novel calorimeter for high-energy cosmic rays in space }
 
\author[a]{ P.W.~Cattaneo, \note{Corresponding author.}}
\author[b,c]{ O.~Adriani,}
\author[d,e]{ S.~Albergo, }
\author[f,g]{L.~Auditore,}
\author[i]{A.~Basti,}
\author[b,c]{E.~Berti,}
\author[h,i]{G.~Bigongiari, }
\author[c]{L.~Bonechi,}
\author[h,i]{S.~Bonechi, }
\author[b,c]{M.~Bongi,}
\author[j]{V.~Bonvicini,}
\author[c]{S.~Bottai,}
\author[h,i]{P.~Brogi, }
\author[k]{G.~Carotenuto, }
\author[c,l]{G.~Castellini, }
\author[b,c]{R.~D'Alessandro,}
\author[c]{S.~Detti,}
\author[m,n]{M.~Fasoli, }
\author[c,o]{N.~Finetti, }
\author[e]{A.~Italiano,}
\author[b,c]{P.~Lenzi,}
\author[h,i]{P.~Maestro, }
\author[h,i]{P.S.~Marrocchesi, }
\author[c]{N.~Mori, }
\author[b,c]{M.~Olmi, }
\author[j]{G.~Orzan, }
\author[b,c,l]{L.~Pacini, }
\author[c]{P.~Papini,}
\author[e]{M.G.~Pellegriti,}
\author[a]{A.~Rappoldi,}
\author[c,l]{S.~Ricciarini}
\author[a]{M.~Rossella,}
\author[e,r]{A.~Sciutto}
\author[b]{P.~Spillantini, }
\author[c]{O.~Starodubtsev,}
\author[h,j]{F.~Stolzi, }
\author[h,i]{J.E.~Suh, }
\author[h,i]{A.~Sulaj, }
\author[b,c]{A.~Tiberio, }
\author[d,e]{A.~Tricomi,}
\author[f,g]{A.~Trifir\`o,}
\author[f,g]{M.~Trimarchi,}
\author[c]{E.~Vannuccini,}
\author[m,n]{A.~Vedda,}
\author[j]{G.~Zampa,}
\author[j]{N.~Zampa,}

\affiliation[a]{INFN Pavia, Via Bassi 6, I-27100 Pavia, Italy}
\affiliation[b]{University of Florence, Department of Physics and Astronomy, via G. Sansone 1, I-50019 Sesto Fiorentino (Firenze), Italy}
\affiliation[c]{INFN Firenze, via B. Rossi 1, I-50019 Sesto Fiorentino (Firenze), Italy}
\affiliation[d]{University of Catania, Department of Physics and Astronomy, via S. Sofia 64, I-95123 Catania, Italy}
\affiliation[e]{INFN Catania, via S. Sofia 64, I-95123 Catania, Italy}
\affiliation[f]{University of Messina, Dipartimento di Scienze MIFT, viale F. Stagno d'~Alcontres, I-98166 Messina, Italy}
\affiliation[g]{INFN Messina, c. da Papardo, sal. Sperone 31, I-98166 S.~Agata, Messina, Italy}
\affiliation[h]{University of Siena, Department of Physical Sciences, Earth and Environment, I-53100 Siena, Italy}
\affiliation[i]{INFN Pisa, via F. Buonarroti 2, I-56127 Pisa, Italy}
\affiliation[j]{INFN Trieste, via Valerio 2, I-34127 Trieste, Italy}
\affiliation[k]{Institute of Polymer, Composites and Biomaterials (CNR), piazzale E. Fermi 1, I-80055 Portici (Napoli), Italy}
\affiliation[l]{IFAC (CNR), via Madonna del Piano 10, I-50019 Sesto Fiorentino (Firenze), Italy}
\affiliation[m]{University of Milano Bicocca, Department of Materials Science, via Cozzi 55, I-20125 Milano, Italy}
\affiliation[n]{INFN Milano Bicocca, piazza della Scienza 3, I-20126 Milano, Italy}
\affiliation[o]{University of L'Aquila, Dipartimento di Scienze Fisiche e Chimiche, Coppito, I-67100 L'Aquila, Italy}
\affiliation[p]{University of Trieste, Dipartimento di Fisica, via Valerio 2, I-34127 Trieste, Italy}
\affiliation[q]{MATIS (CNR), via S. Sofia 64, I-95123 Catania, Italy}
\affiliation[r]{IMM (CNR), Ottava strada 5, I-95121 Catania, Italy}

\emailAdd{paolo.cattaneo@pv.infn.it}

\abstract{
In order to extend the direct observation of high-energy cosmic rays up to the PeV region, 
highly performing calorimeters with large geometrical acceptance and high energy resolution
are required. Within the constraint of the total mass of the apparatus, crucial for a space mission, 
the calorimeters must be optimized with respect to their geometrical acceptance, granularity 
and absorption depth.
CaloCube is a homogeneous calorimeter with cubic geometry, to maximise the acceptance being sensitive to
particles from every direction in space; granularity is obtained by relying 
on small cubic scintillating crystals as active elements.
Different scintillating materials have been studied. The crystal sizes and spacing among them have been optimized 
with respect to the energy resolution. 
A prototype, based on CsI(Tl) cubic crystals, has been constructed and tested with particle beams. 
Some results of tests with different beams at CERN are presented.
}

\proceeding{INSTR17: Instrumentation for Colliding Beam Physics \\
  27 February -- 3 March, 2017 \\
  Novosibirsk, Russia}

\begin{document}

\maketitle
%
\section{Introduction}
The scientific rational driving CaloCube\footnotemark \footnotetext{CaloCube is an R\&D project financed by INFN for 3+1 years (end 2017).}
\cite{D'Alessandro:2016ylh} requirements is the direct measurement of the charges cosmic
rays close to the 'knee', the PeV region where the inclusive spectrum of cosmic rays changes slope 
becoming steeper and the composition progressively heavier. So far this region has been accessible
only with ground based experiments detecting induced air-showers. This indirect measurements have 
large acceptance but result in large systematic errors due to model dependence for what concerns 
energy resolution and composition studies. 
Direct CR detection overcomes those limitations, but suffers from limited exposure, due to constraint 
in the apparatus mass in space mission. This constraint have limited the measurement of H and He spectra
to 100 TeV and heavier nuclei to no more than 1 TeV. 
The detection of the knees of the H and He spectra requires an acceptance of at least 2.5 m$^2$ sr$\times 5$ yr
while the energy resolution is less constrained, being sufficient $\sim$ 40\%. 
A possible detector design satisfying these requirements is a charge measuring device followed by a calorimeter.

Additionally this calorimeter can measure the cosmic rays electrons and,
if coupled to a tracker-converter system and an anticoincidence shield, the high-energy gamma radiation.
This possibility requires an energy resolution better than 2\%\ and a high h/e rejection factor above 1 TeV.

\section{The CaloCube concept}

The goal of CaloCube 
is to achieve the above discussed performance with a
space-borne calorimeter, within the constraint
dictated by the limitation in weight ($\sim 2$ tons).

\begin{figure}[ht]
  \centering
   \includegraphics[width=0.95\textwidth]{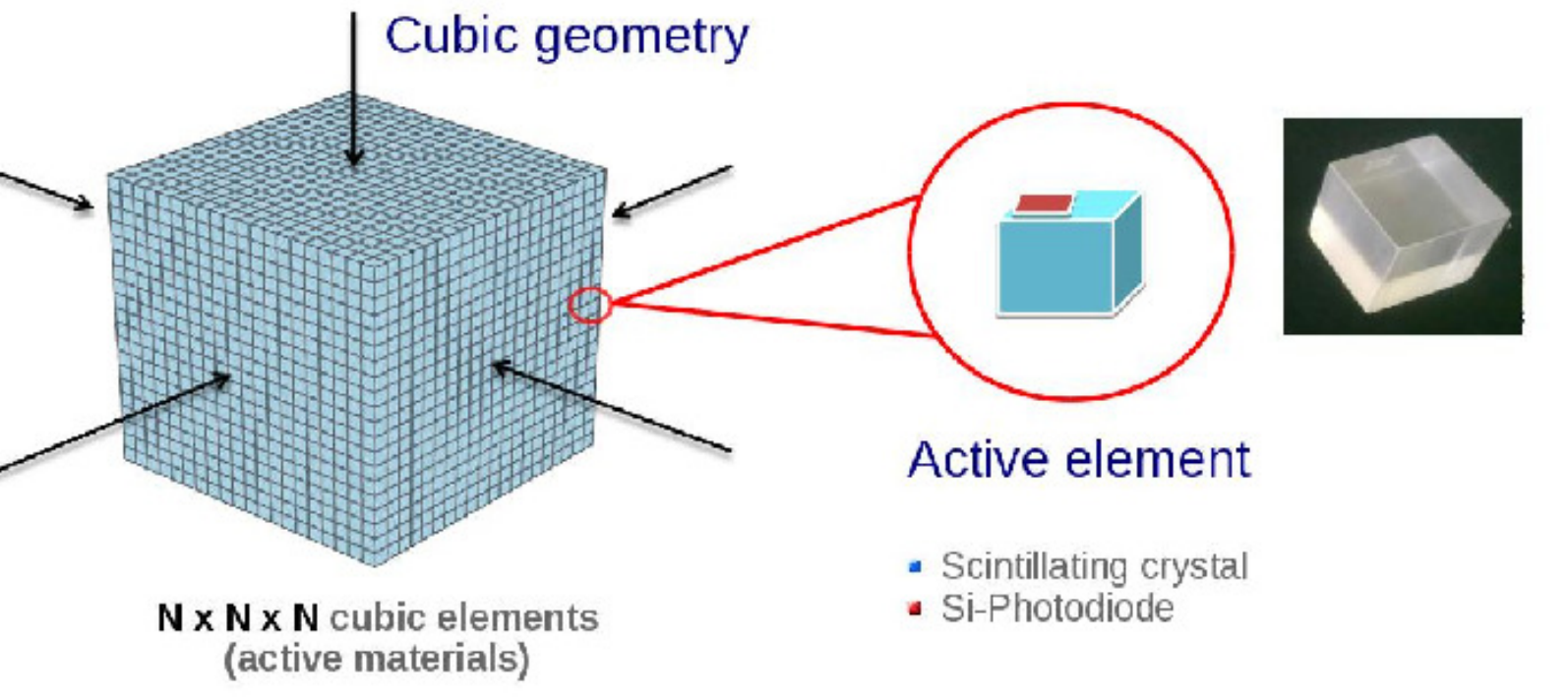}
   \caption{\small The CaloCube layout is based on a quasi-isotropic geometry with cubic symmetry,
             where the active volume is made of NxNxN small cubic scintillator crystals,
             optically isolated from each other, read from one or more solid state detectors
             (Si-Photodiode).}
\label{Cube}
\end{figure}

The proposed solution consists in a 3D array of cubic scintillating crystals, 
optically isolated from each other, readout by one or more photodiodes (PDs), arranged
to form a cube (see Fig.~\ref{Cube}). The homogeneous detector geometry 
provides the possibility to collect particles from five faces out of six 
(bottom is excluded), so the geometrical acceptance for a fixed mass budget
is maximised. The cubic crystals, acting as active absorber, provides good
energy resolution, while the high granularity provides 3D shower imaging,
providing information for leakage correction and h/e separation.

Such stringent requirements can be obtained using a calorimeter in conjunction with a 
dE/dx detector, and for this purpose the CaloCube project was created,
with the aim of designing and optimising of a calorimeter for measurements of
high-energy cosmic rays in space \cite{Bongi:2015gma, Vannuccini:2017gsa}.
%
\section{The Monte Carlo simulations and the expected performances}
\label{sec:mc}

The calorimeter response depends on the adopted geometry and on the material used for
the single scintillating crystals. For this reason, an accurate Monte Carlo,
based on the FLUKA package, has been developed to carefully study the calorimeter
operation in different configurations.
 
A NxNxN cubic geometry, with cubes of about 1 Moli\`ere radius size each,
has been studied with different scintillating materials, sizes and gaps (distance between the
adjacent cubes). Different configurations, comparable in terms of total weight ($W_{tot} \simeq 2$~tons)
and active volume fraction (about 78\%) are shown  in Table~\ref{configurations}.
\begin{table}[h]
\centering
\footnotesize {
\begin{tabular}{|c|c|c|c|c|c|}
\hline
        &  {\bf CsI(Tl)} & {\bf BaF$_\mathbf{2}$} & {\bf YAP(Yb)} & {\bf BGO} & {\bf LYSO(Ce)} \\
\hline
{\it l (cm)} &  3.60           & 3.20                   & 2.40          & 2.30      & 2.10 \\
{\it Gap (cm)} & 0.30          & 0.27                   & 0.20          & 0.19      & 0.18 \\
{\it N cubes}  & 20 x 20 x 20  & 22 x 22 x 22           & 28 x 28 x 28  & 27 x 27 x 27 & 30 x 30 x 30 \\
{\it L (cm)}  & 78.0           & 76.3                   & 72.8          & 67.2         & 68.2 \\
$L_{int}$ $(\lambda_{int}$) & 1.80 & 2.31                  & 3.09          & 2.72         & 3.01 \\
$L_{rad}$ $(X_{0})$ & 38.88 & 34.73                       & 24.96         & 55.54        & 53.75 \\
$GF$ $(m^2 sr)$     & 9.56  & 9.15                        & 8.32          & 7.10         & 7.35 \\
\hline
\end{tabular}
}
\caption{\small Main parameters of the simulated calorimeter geometries (see Sect.~\ref{sec:mc} for explanation).}
\label{configurations}
\end{table}
Among the five materials, LYSO (Lu$_{1.8}$Y$_{0.2}$SiO$_5$(Ce)) is the best one
for protons, due to the better shower containment, which compensates for the
smaller volume (due to its high density of 7.1 g/cm$^3$).

The collection efficiency of the scintillation light depends on the type of
coating used to reflect the light and isolate each crystal from the adjacent
ones. Several different materials have been tested measuring the signal induced
by a 5.5 MeV $\alpha$ emitted by a $^{241}$Am source, and the results (in terms
of signal amplitude) are shown in Fig.~\ref{Coating}.
%

\section{The prototype }

\begin{figure}[h]
\centering
  \begin{minipage}[b]{0.47\textwidth}
  \includegraphics[width=1.0\textwidth]{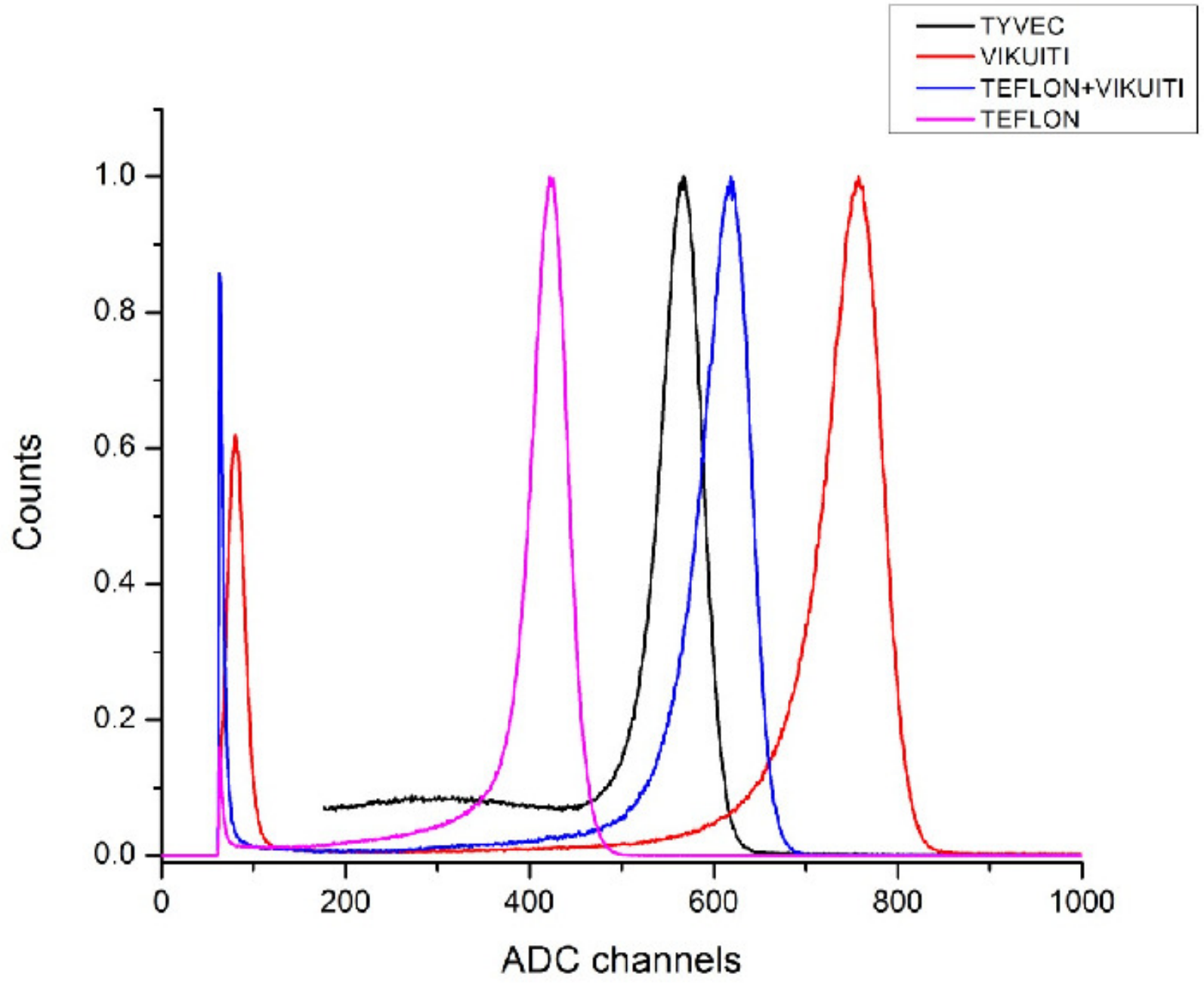}
  \vfill
  \caption{\small Signal amplitude due to 5.5 MeV $\alpha$ particles of a single CsI(Tl) scintillator crystal
           (36 mm)$^2$ with different coating materials.}
  \label{Coating}
  \end{minipage}
  \hfill
  \begin{minipage}[b]{0.47\textwidth}
  \centering
  \includegraphics[width=1.0\textwidth]{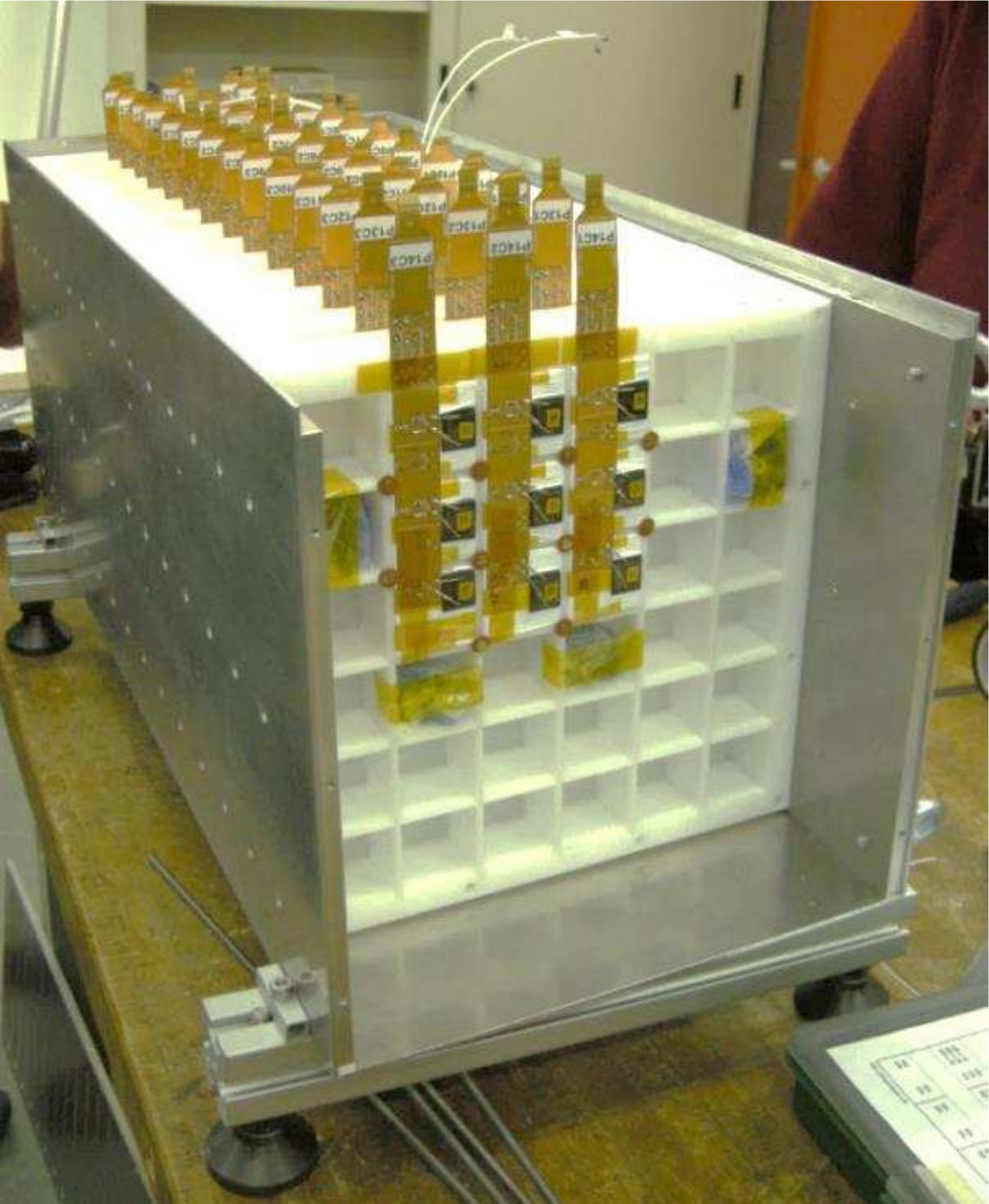}
  \caption{\small The prototype with 3 x 3 x 15 CsI(Tl) crystals tested under
           beam at CERN in 2015.}
  \label{Prototype}
  \end{minipage}
\end{figure}

A small mechanical structure has been built, to allow the test of different
geometric configurations on an accelerator beam.
A prototype with 135 CsI(Tl) cubic crystals of 3.6 cm size, arranged in 15 planes of 3~x~3
cubes each, with a gap of 0.4 cm between them, has been built (Fig.~\ref{Prototype}) 
and tested on accelerator beams. 

This prototype has a shower containment of about 1.5 Moli\`ere
radius ($R_M$) and an active depth of 28.4 $X_0$ and 1.35 $\lambda_{\mathrm int}$,
and has been tested at CERN with different particle beams, summarised in Table~\ref{testbeam}.
Each crystal is wrapped with few layers of Teflon tape and optically
coupled to a single PD, VTH2090H from Excelitas, a large-area
($\sim 100$~mm$^2$) sensor that allows to clearly detect minimum-ionising protons with a
signal-to-noise ratio of $\sim 15$. 
One of the most challenging requirements is the very large dynamic range ($10^7$)
ranging from 20 MeV for minimum ionising protons to 10\%\ of the energy of a PeV proton.
This will be accomplished by using also a second small area PD ($\sim 1$~mm$^2$).
The front-end electronics is based on a custom designed high dynamic-range, 
low-noise ASIC, characterised by a dynamic range of 52.2 pC and an ENC$\sim 0.5$ fC 
at 70 pF input capacitance.

\section{Beam test and results}
\begin{table}[ht]
\centering
\footnotesize {
\begin{tabular}{|l|l|l|}
\hline
{\bf Test}     & {\bf Beam}      & {\bf Energy} \\
\hline
Feb. 2013      & ions from Pb + Be\, A/Z=2   &  13-30 GeV \\
Mar. 2015      & Ions from Ar + Poly   &  19-30 GeV \\
Aug./Sep. 2015 & $\mu$, $\pi$, e &  50, 75, 150, 180 GeV \\
\hline
\end{tabular}
}
\caption{Parameters of the beam tests.}
\label{testbeam}
\end{table}

\begin{figure}[ht]
\centering
  \begin{minipage}[b]{0.47\textwidth}
  \includegraphics[width=1.0\textwidth]{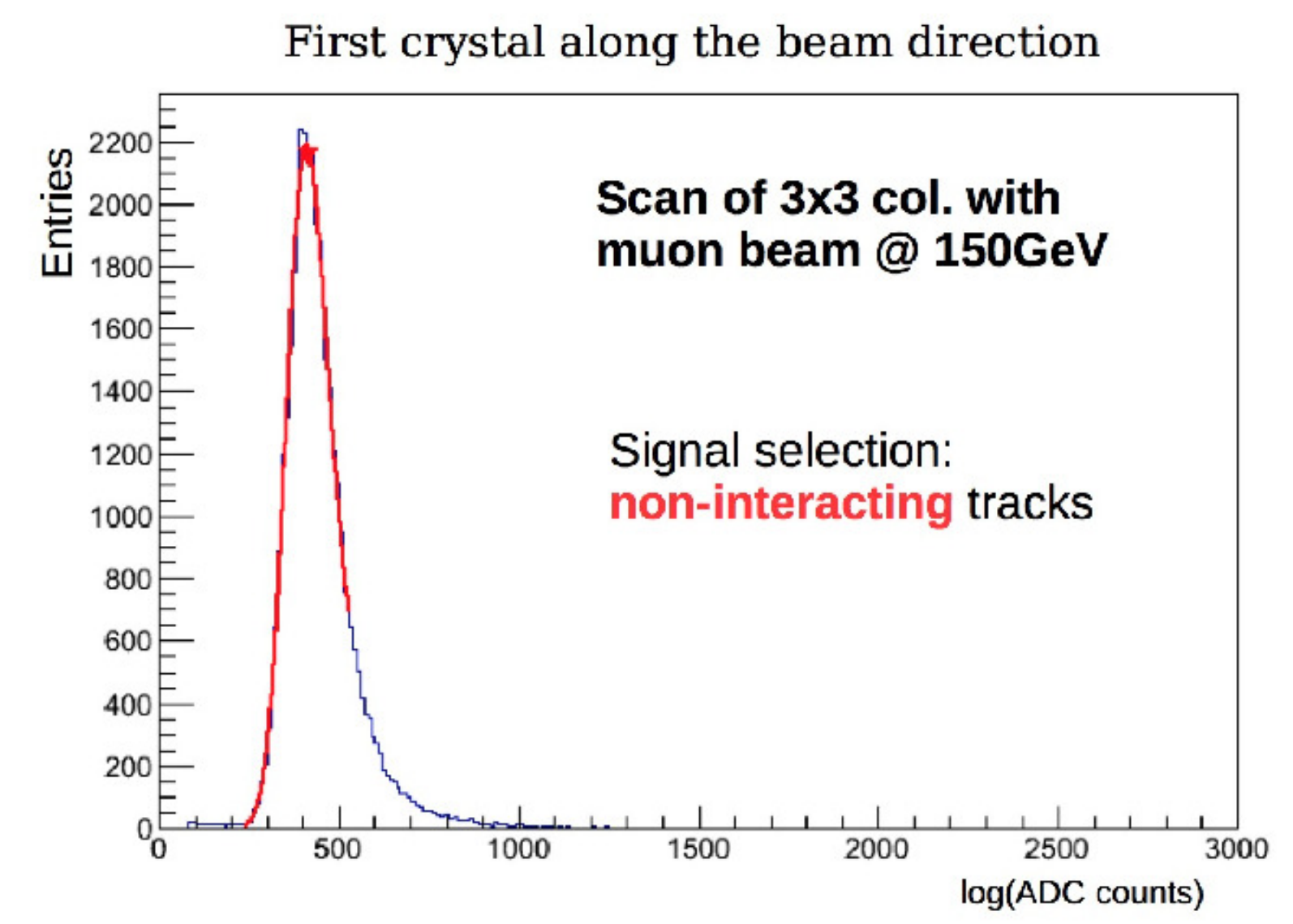}
  \vfill
  \caption{\small Response of a single crystal to a non-interacting track at 150 GeV (mainly $\mu$).
          The distribution is well fit with a Landau distribution (red).}
  \label{Calibration}
  \end{minipage}
  \hfill
  \begin{minipage}[b]{0.47\textwidth}
  \centering
  \includegraphics[width=1.0\textwidth]{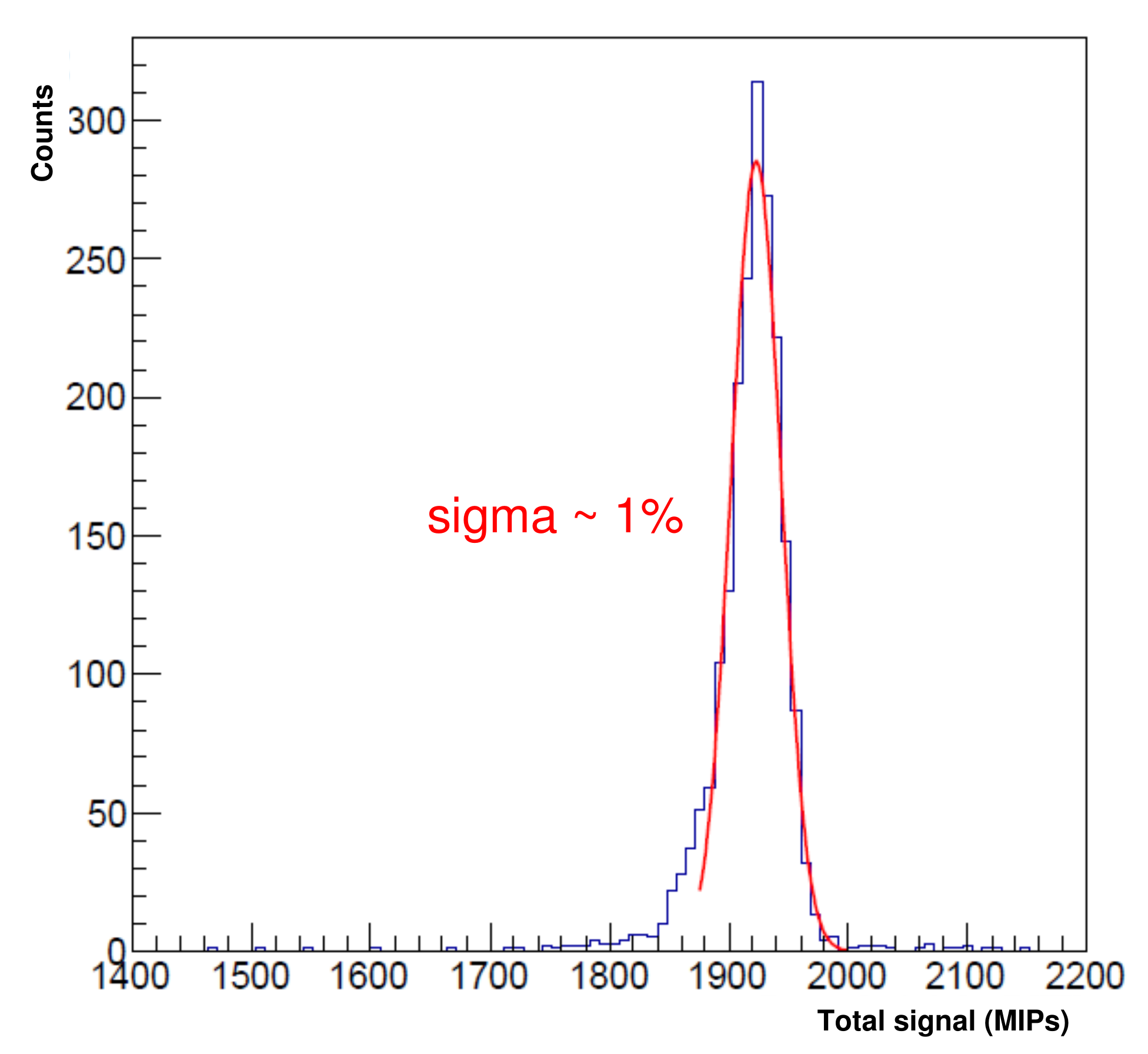}
  \caption{\small Measured distribution of total energy (expressed in m.i.p.) released with a 
           50 GeV electrons beam, fit with with the expected distribution (red).
           The resulting energy resolution is about 1\%.}
  \label{Electrons}
  \end{minipage}
\end{figure}

During the beam test at CERN in summer 2015, the prototype was
exposed to $\mu^\pm$ beams, to obtain the response of each
crystal to a minimum ionising particles (m.i.p.), thus determining the individual
conversion factors between the deposited energy and the photodiode signal
(Fig.~\ref{Calibration}). These factors were later used to equalise the response of
all the cubes.

Then an estimate of the energy resolution has been determined, exposing the
calorimeter to $\mathrm{e^\pm}$ beams of different energy, and determining the total
deposited energy, given by the sum of the equalised signals of all the cubes.
A preliminary result is shown in Fig.~\ref{Electrons}, referring to a beam
of 50 GeV electrons. In this case the measured total energy (expressed in m.i.p)
is in good agreement with the expectation, and the corresponding resolution
is at level of 1\%.
In the 2013 beam test the prototype was exposed to ion beams of
12.8 and 30 GeV/n, containing A/Z=2 fragments produced by a primary Pb beam
colliding with a Be target.

\begin{figure}[ht]
\centering
  \begin{minipage}[b]{0.47\textwidth}
  \includegraphics[width=1.00\textwidth]{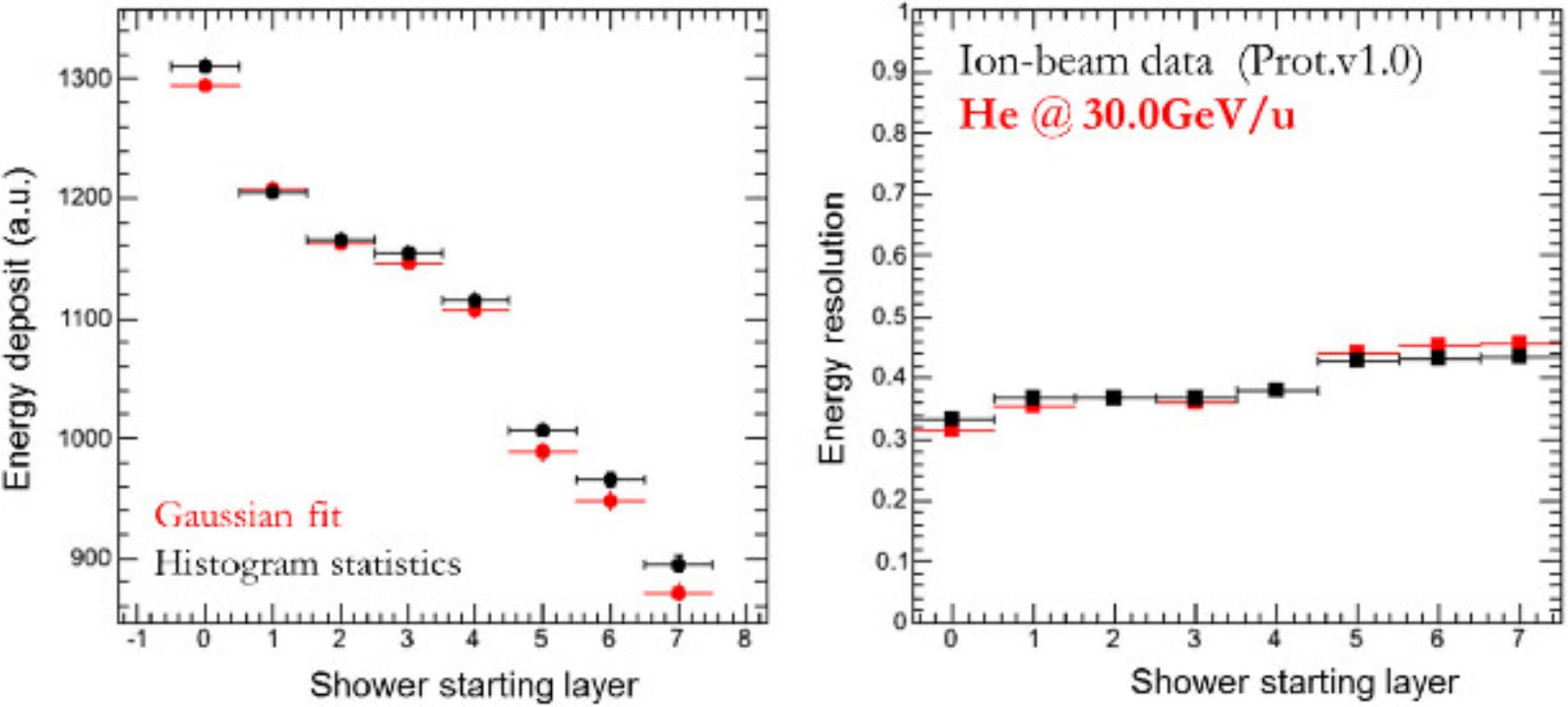}
  \vfill
  \caption{\small Energy deposit (left), in MIP, and energy resolution (right) measured for He nuclei, as a function of the layer where the shower starts.}
  \label{Layers}
  \end{minipage}
  \hfill
  \begin{minipage}[b]{0.47\textwidth}
  \centering
  \includegraphics[width=1.00\textwidth]{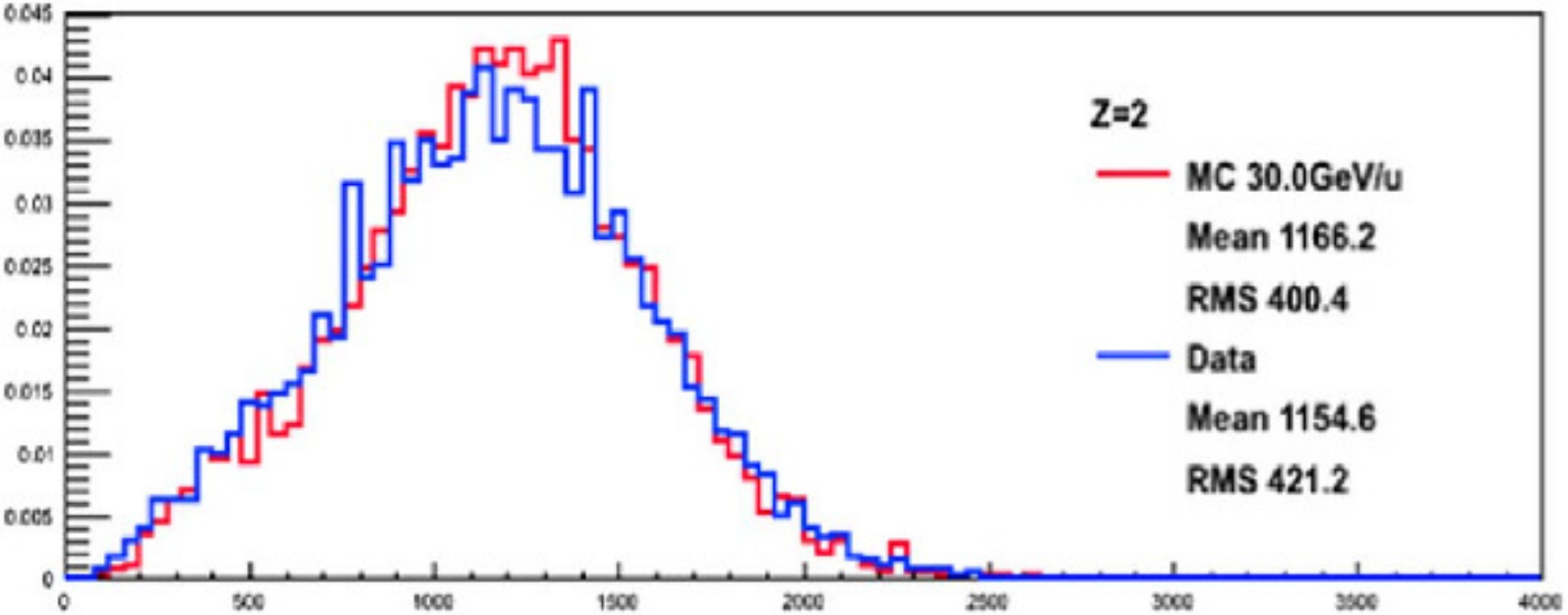}
  \caption{Distribution of the energy deposit (in MIP) of 30 GeV/n He ions for data (blue) and MC (red).}
  \label{Ions}
  \end{minipage}
\end{figure}

The single-crystal performances were studied, by selecting non-interacting ions. 
The observed dispersion among crystal responses was
about 15\%, with an average signal-to-noise ratio for deuterons of $\sim~10$.
The single-crystal responses were equalised by normalising to the energy
deposit of non-interacting He nuclei (set by definition to 4 MIP), the
most abundant fragments.
Showers were classified on the basis of the starting point, which in the beam-test
set-up univocally determines the shower containment. Fig.~\ref{Layers} shows the
energy deposit (left) and its relative fluctuations (right) for He induced
showers, versus the shower starting layer; in spite of the
significant leakage for showers starting progressively deeper inside the
calorimeter, the energy resolution is almost constant and better than
40\%\ down to the fifth layer.

A Monte Carlo model of the prototype has been developed in Fluka and its
predicted response is shown in Fig.~\ref{Ions} in
comparison with real data. A fine tuning of the Monte Carlo was
necessary in order to reproduce the beam-test data. In particular, an
additional spread of 4.5\% on the single-crystal responses and an optical
cross-talk of 14\%\ were introduced.
%
%

\acknowledgments

This work was supported by the Istituto Nazionale di Fisica
Nucleare (INFN), through the CaloCube project, and by the H2020
project AIDA-2020, GA no. 654168. The authors thank CERN for the
allocation of beam time at the North Area test facility.


\bibliographystyle{JHEP}
\bibliography{CaloCube}

\providecommand{\href}[2]{#2}\begingroup\raggedright\begin{thebibliography}{1}

\bibitem{D'Alessandro:2016ylh}
R.~D'Alessandro et~al., \emph{{Calocube: A highly segmented calorimeter for a
  space based experiment}},
  \href{http://dx.doi.org/10.1016/j.nima.2015.09.073}{\emph{Nucl. Instrum.
  Meth.} {\bfseries A824} (2016) 609--613}.

\bibitem{Bongi:2015gma}
M.~Bongi et~al., \emph{{CALOCUBE: an approach to high-granularity and
  homogenous calorimetry for space based detectors}},
  \href{http://dx.doi.org/10.1088/1742-6596/587/1/012029}{\emph{J. Phys. Conf.
  Ser.} {\bfseries 587} (2015) 012029}.

\bibitem{Vannuccini:2017gsa}
E.~Vannuccini et~al., \emph{{CaloCube: A new-concept calorimeter for the
  detection of high-energy cosmic rays in space}},
  \href{http://dx.doi.org/10.1016/j.nima.2016.07.014}{\emph{Nucl. Instrum.
  Meth.} {\bfseries A845} (2017) 421--424}.

\end{thebibliography}\endgroup
\end{document}